\documentclass[12pt]{article}
\usepackage[english]{babel}
\usepackage{amssymb,graphicx,wrapfig}
\makeindex

\begin{document}

\vspace*{1.2cm}

\thispagestyle{empty}

\begin{center}

{\LARGE \bf Odderon: Lost or/and Found?}

\par\vspace*{7mm}\par

{

\bigskip

\large \bf Vladimir Petrov\footnote{Speaker}  and Nikolay Tkachenko }

\bigskip

{\large \bf  E-Mail: Vladimir.Petrov@ihep.ru}

\bigskip

{Logunov Institute for High Energy Physics, NRC "Kurchatov Institute", Protvino, RF}

\bigskip

{\it Presented at the Low-$x$ Workshop, Elba Island, Italy, September 27--October 1 2021}

\vspace*{15mm}

\end{center}

\vspace*{1mm}

\begin{abstract}
   This is a quick survey of theoretical and experimental efforts to
   understand and identify the Odderon.
\end{abstract}

\section{Introduction}

 In a long and rich history of the studies on high energy hadron
 interactions two kinds of strong ("nuclear") forces are constantly
 featured: C-even and C-odd ones. 

 C-even forces represented by the Pomeron and f-Reggeon bear a universal
 character, i.e. they are like the forces of gravity acting as the
 universal attraction of all hadrons to each other. On the contrary,
 C-odd forces, which in high-energy physics are associated with
 $\rho$-,$\omega$-, etc. Reggeons resemble electromagnetic interactions,
 when charges of the same sign (particle-particle) repel, and charges
 of different signs (particle-antiparticle) are attracted to each other.
 
Until the beginning of the 70s, the belief reigned that with the increase
in collision energy, the main C-even agent, the Pomeron, plays an increasingly
dominant role, while the C-odd forces become less and less significant
("die out") and finally can be neglected.

Such a paradigm was challenged in the works of B. Nicolescu et al.\cite{luk},
in which a new notion was introduced, later dubbed "Odderon", which, being
a part of the amplitude subleading (w.r.t.the Pomeron) in the imaginary part
of the scattering amplitude, becomes the leading one in the real part.

It worth noticing that the very term" Odderon" looks akin to the name of
some new Reggeon but according to \cite{luk} this was but a specific contribution
to the C-odd part of the scattering amplitude. No Reggeon (or a particle)
was associated with this "Odderon"\footnote{Later the Odderon option as formulated
in \cite{luk} was dubbed (after serious modfifications) the "Maximal Odderon"
in contrast to other Odderon incarnations (e.g. as a C-odd Reggeon). As we
will not concern these models , we will  use the term "Odderon" in the sense
of "Maximal Odderon" as well.}.

Below we will try to trace almost a half-of-century history of the Odderon
concept and attempts and efforts to find its experimental manifestations.

\newpage

\section{Odderon: Predictions and Nature}

Which observables are potentially suitable for identifying the possible existence
of the Odderon?  Below we briefly describe a few options used.

\subsection{Difference of the proton-proton  and anti proton-proton total
cross-sections.}

\begin{wrapfigure}[40]{l}{120mm}
\vspace{-6.1mm}
\includegraphics[width=120mm]{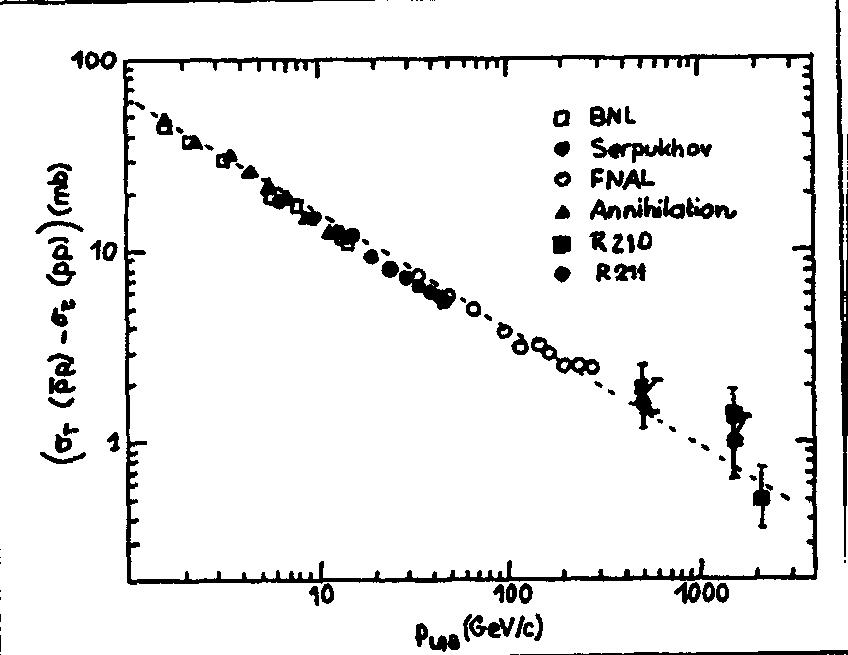}
\vspace{-7.6mm}
\caption{The energy evolution of the difference $ \Delta\sigma =
\sigma_{tot}^{\bar{p} p} - \sigma_{tot}^{pp}$ at $ \sqrt{s}
\leqslant 60\: GeV $.}
\label{p1}
\vspace{0.1mm}
\includegraphics[width=120mm]{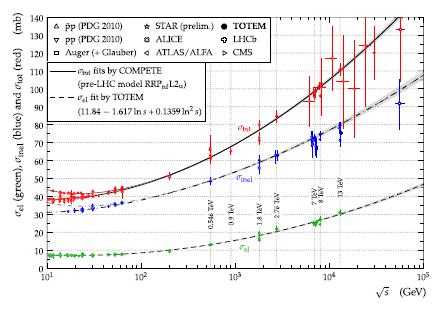}
\vspace{-9.6mm}
\caption{The energy evolution of the cross-sections in  $\bar{p}p$ and
$pp$ collisions.}
\label{p2}
\end{wrapfigure}

The simplest one is the difference between the total cross-sections of
$\bar{p}p$ and $pp$ interactions 
$$\Delta\sigma = \sigma_{\mbox{tot}}^{\bar{p}p} - \sigma_{\mbox{tot}}^{pp}.$$
because it is exactly a C-odd quantity.

The first paper in Ref.\cite{luk} predicted that "at high energies
($\sqrt{s}$)" 

\begin{center}
$\mid \Delta\sigma \mid \sim \ln s$
\end{center}
i.e. grows indefinitely with energy. 
Pre-ISR data showed that $\Delta\sigma$ is positive and decreases with energy
growth.

The ISR data ($\sqrt{s}= 20\div 63\:\mbox{ GeV}$) confirmed this trend and gave
the last opportunity to compare $\sigma_{\mbox{tot}}^{\bar{p}p} $ and
$\sigma_{\mbox{tot}}^{pp}$ at the same energy. The minimum value of the difference
as measured at the ISR \cite {amb} was

$$\Delta\sigma (52.8\mbox{ GeV}) = 1.49\pm 0.35\mbox{ mb}$$

Fig.1 \cite{Ow}  shows the early result for $\Delta\sigma$ for laboratory
energies

$$E_{\mbox{\mbox{lab}}}\approx p_{\mbox{\small{lab}}}c \leqslant
2000\mbox{ GeV}$$
(cms energy up to $ 60\mbox{ GeV}$). 

However, in the mentioned first article on the Odderon it was argued that $\Delta\sigma $ should drop till

 $\sqrt{s} \approx 24$ GeV

where it should disappear and then (after it would turn negative) would begin
to indefinitely grow in absolute value achieving
$ - \:\mathcal{O}( 10\mbox{ mb})$
at $p_{\mbox{lab}}c = 10^{4}\mbox{ GeV}~ (\sqrt{s} \approx
150\mbox{ GeV}$). This would be a clear evidence in favour of the Odderon
as formulated in \cite{luk}  but the ISR measurements, as we see,  ruled
out such an option.

Meanwhile $\Delta\sigma$ well collaborated with the prediction of the
Regge pole scheme
$$\Delta\sigma \sim s^{\alpha_{-}(0)-1}$$
where $\alpha_{-}(0)$ is the intercept of the "secondary" Reggeons
($\rho,~\omega$ etc). As generically $\alpha_{-}(0)\approx 0.5$
we see that Fig.1 seemed to confirm the asymptotic disappearing of
$\Delta\sigma$.

This , however, did not discourage the Odderon proponents who argued that
the crossover of $\sigma_{\mbox{tot}}^{\bar{p} p}$ and
$\sigma_{\mbox{tot}}^{pp}$ had a chance to show up  at higher energies.

\begin{wrapfigure}[20]{l}{120mm}
\vspace{-0.1mm}
\includegraphics[width=120mm]{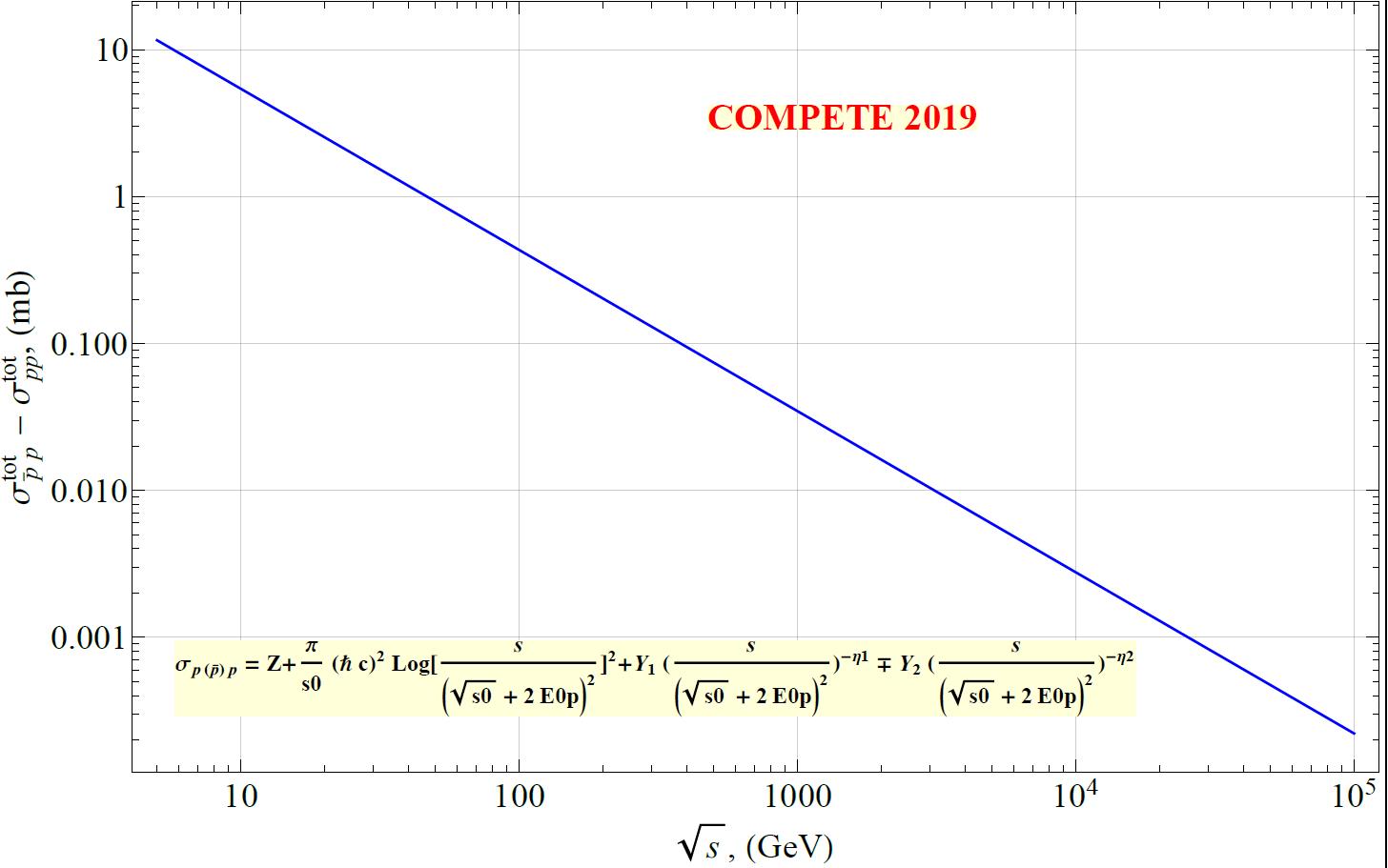}
\vspace{-7.6mm}
\caption{The energy evolution of $ \Delta\sigma $ as given by the
COMPETE parametrization.}
\label{p3}
\end{wrapfigure}

Postponing the $ Sp\bar{p}S $ results for a bit later, let us  come to the
highest energies achieved by now, i.e. 2 TeV for $ \bar{p} p$ and 13 TeV
for $pp$. A straightforward comparison between the two channels is still
impossible because of the absence of the relevant data at the same energy,
so we take the COMPETE parametrization \cite{ed} which describes the data
on $\sigma_{tot}^{\bar{p}p}$ and $\sigma_{\mbox{tot}}^{pp}$ very well.
This is pictured in Fig.2 \cite{Ant2} .

The COMPETE predicts for $\Delta\sigma$ the stable decrease  as is seen
in Fig.3.

So, it seems that the difference in total cross sections is not the best place to look for manifestations of the Odderon\footnote{The enthusiasts of the "Maximal Odderon" still insist that the cross-over will occur though such claims already do not look very convincing} . However, this could only mean that the Odderon does not couple significantly to the imaginary part of the \textit{forward} scattering amplitude while a possibility of a noticeable coupling  to the real part of the forward scattering amplitude is quite conceivable.

\subsection{Early sounding of the Odderon via ReF(s,0)/ImF(s,0).}

So, in addition to the difference between the cross sections,
the quantity\vspace{-2.1mm}
$$\rho = \mbox{Re}F(s,0)/\mbox{Im}F(s,0).\vspace{-1.1mm}$$

($F(s,t)$ stands for the elastic scattering amplitude) seemed to be  a suitable
observable quite accessible at the $ Sp\bar{p}S $ collider. Although there
were no corresponding $pp$ data, the Odderon contribution could manifest
itself in $\rho^{\bar{p}p}$ through the dispersion relations, as a sort of
"echo" from the u-channel. 

During the functioning of the $Sp\bar{p}S$, two dedicated measurements were made in the UA4 and UA4/2 experiments. 

The result obtained in UA4 \cite{ua4/1}  became a sensation: instead of the expected value of about $ 0.10\div 0.15 $, it turned out that\vspace{-2.1mm}
$$\rho^{\bar{p}p}~(\mbox{UA4}) = 0.24 \pm 0.04 !$$
The number caused a flow of publications, often containing the most fantastic scenarios, but the enthusiasts of the "Maximal Odderon" felt themselves to be the main beneficiaries \cite{nic2} . It was the notorious "maximality" that seemed to be the reason for such a large value of $ \rho $.

Six years passed in discussions, conferences , talks and articles till a new sensation broke out. A "remeasurement" undertaken by the UA4 collaboration (under the nickname UA4/2) produced the following result \cite{ua4/2}:\vspace{-1.1mm}
$$\rho^{\bar{p}p}~ (\mbox{UA4}/2) = 0.135 \pm 0.015 .$$
Concerning the 1987 result it was said: "{\it The previous result
$\rho=0.24+0.04$ obtained with a poor beam optics, a factor eleven less
statistics and much less control of systematic effects should be considered
as superseded.}"\cite{ua4/2}.

So the "Maximal Odderon" was again not lucky: after several years of triumph, disappointment came. 

But ahead there  was a new take-off, although one had to wait a long time, almost a quarter of a century. 

\subsection{A resurrection of the "Maximal Odderon" or...?}

In December 2017, one of us (V.P.) had a long discussion with S. Giani and J. Ka\v{s}par about their just obtained result on extracting the value of $ \rho $ from the data of the TOTEM collaboration on the differential cross section for elastic proton-proton scattering in the region of Coulomb-nuclear interference at 13 TeV.

The point was that this value  ($ \rho = 0.1 $) essentially coincided with the value of $ \rho $ obtained earlier in the theoretical

article by B. Nicolescu and E. Martynov, in which an attempt was made to describe a large amount of data in the framework of a highly modified version of the "Maximum Odderon" model.

On this basis, the conclusion was made: a new particle was discovered, the "Odderon"consisting of 3 gluons! 

The news attracted attention of the public media. For instance, a few months later an article appeared in  "The Newsweek" under the title:

"What's an Odderon and Did CERN Just Revealed it Exists?" The more professional CERN Courier placed in March 2018 an article " Oddball Antics in pp Collisions" and finally in Match 2021 "Odderon Discovered" with a statement:

"The TOTEM collaboration at the LHC, in collaboration with the DØ collaboration at the former Tevatron collider at Fermilab, have announced the discovery of the Odderon – an elusive three-gluon state predicted almost 50 years ago."

Without a doubt, the discovery of a new particle is a great event and an outstanding achievement for any experiment.

Let us now look at two publications concerning these findings.

The first one appeared in 2019 \cite{TOT1} and was devoted to "probing

the existence of a colourless C-odd three-gluon compound state" on the basis of the retrieval of the parameter $ \rho $ from the data on elastic proton-proton scattering in the region of Coulomb-nuclear interference at 13 TeV. The second one will be commented in the next subsection.

As was mentioned above, the conclusion about the discovery  of the "C-odd three-gluon compound state" was made because of coincidence of the measured value of $ \rho $ with the approximately the same value appearing in the model of "Maximal Odderon" \cite{nic3}. 

What is interesting, in the model suggested in \cite{nic3} the  authors do not deal with "gluons" at all ( because their arguments do not use QCD) and define the Odderon as follows:

"The Odderon is defined as a singularity in the complex $j$-plane,
located at $ j=1 $ when $ t=0 $ and which contributes to the odd-
under-crossing amplitude $ F_{-} $". 

However, the crossing-odd amplitude (negative signature)with  $j=1$
cannot have singularity at $t=0$ because this is the \textit{physical}
p-wave partial amplitude in $\bar{p}p$ channel. Otherwise the axiomatic
bounds (assuming non-zero mass gaps in any channel)would be violated while
the authors use precisely these bounds to justify their "maximum" choice
for the C-odd amplitude. Otherwise, we would be forced to assume that there
is no color confinement. This in turn would naturally give rise to an infrared
singularity at $t=0$. I do not believe that authors of \cite{nic3} meant
such a radical scenario. Although, in this case, gluons would appear, indeed,
but in an amount significantly exceeding 3. 

In contrast, the C-even (positive signature) amplitude $F_{+}(1,t)$  may
well have a singularity at $ t=0 $ , since for it the value $j=1$ is not
physical.

A detailed criticism of the "Maximal Odderon model" in both conceptual and
descriptive aspects can be found in Ref.\cite{ptr1}.

There is one more aspect of this topic that I would like to touch upon. 

In the article \cite{TOT1} some phenomenological model for the strong
interaction amplitude was used for description of the data and hence, for
retrieving parameters, e.g.  $ \rho $. However, this model does not exhibit
the Odderon singularity as does the strong interaction amplitude described
in \cite{nic3}.

So, the coincidence of the values of $ \rho $ seems accidental, not related with the presence or absence of the Odderon singularity as assumed in \cite{TOT1}.

Our conclusion from this reasoning is that no specific value of the parameter
$\rho$ can be considered as an evidence of presence or absence of the Odderon.

Being the $\rho$-parameter of the "forward origin" this is in line with the
above conclusion about another forward observable $\Delta\sigma$ and implies
that the Odderon, if exists, should be probed at non-zero transferred momenta.

Its decoupling at $t=0$ is evidently related with absence of massless states
in the p-wave partial amplitude $ F_{-}(1,t) $in the $\bar{p}p$-channel,
i.e. actually with confinement.\vspace{-2.1mm} 

\subsection{A step aside: the Odderon at nonzero t. An old friend? }

\begin{wrapfigure}[26]{l}{95mm}
\vspace{-0.1mm}
\includegraphics[width=95mm]{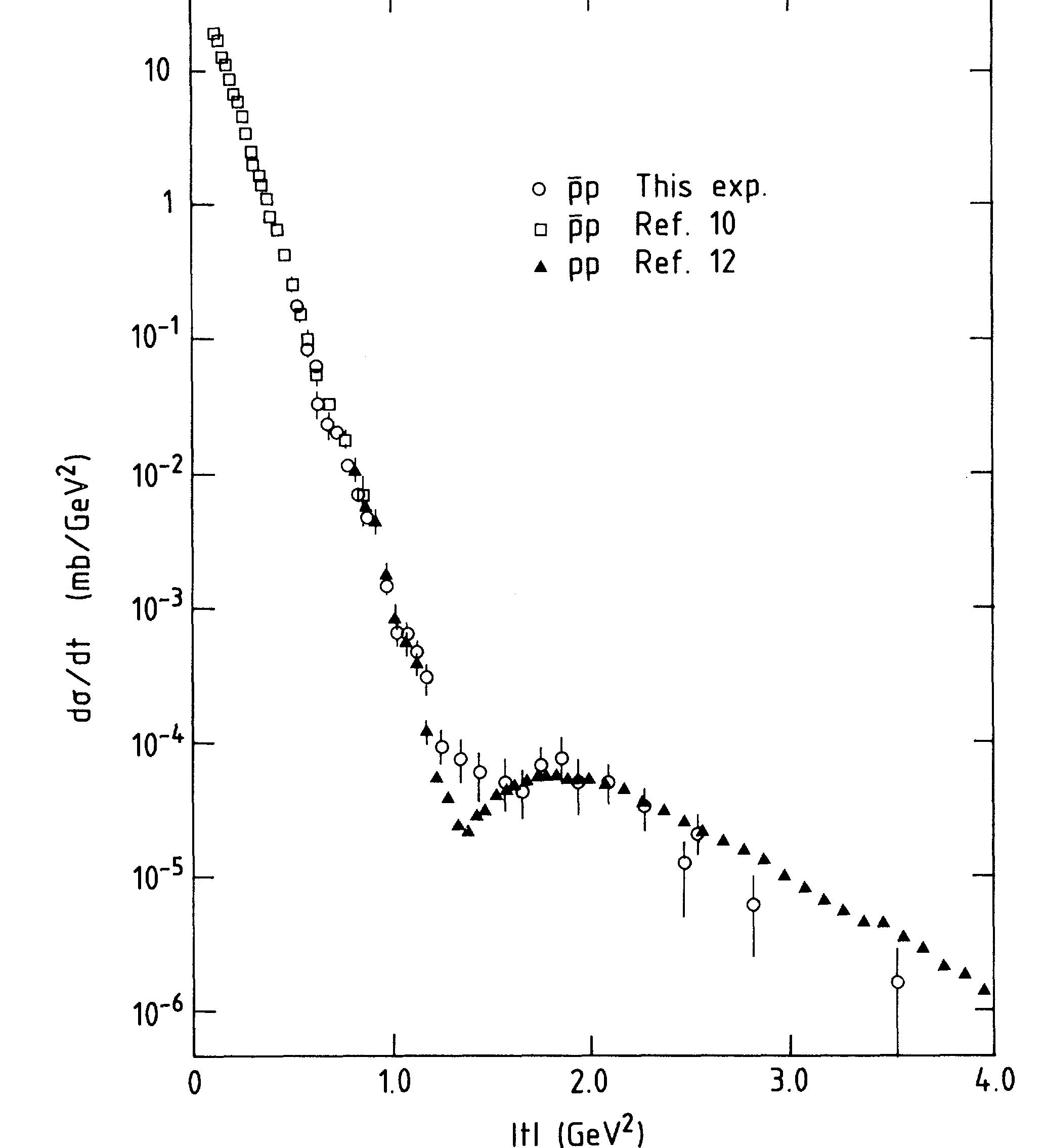}
\vspace{-4.6mm}
\caption{The comparison of ($ pp $) and  $ \bar{p}p $ cross-sections.}
\label{p4}
\end{wrapfigure}

In no way all above said means that the Odderon does not exist or is unobservable. We argued only about forward observables. If $ t\neq 0 $ the only way to search for it is a comparison of differential cross-sections in $ pp $ and $ \bar{p}p $-channels.

And here we cannot help but recall the good old ISR. We find that as early as in 1985 a dedicated measurements were made \cite{isrod}  to compare  the differential cross-section of elastic $ pp $ and $ \bar{p}p $ scattering at the same energy ($ \sqrt{s}= 53 $ GeV).

Fig.4  \cite{isrod} shows that the two cross-sections almost coincide except the vicinity of the dip ($ pp $) and shoulder ( $ \bar{p}p $).

Fig.4. The comparison of ($ pp $) and  $ \bar{p}p $ cross-sections.

Fig.5 shows the ratio of the two cross-sections which differs from 1 only at $ t $ in the vicinity of dip/shoulder.

It is clear that the cause of the difference is a C-odd force. But which one? Is it the manifestation of the well known secondary Reggeons which are responsible for a non zero  

$ \Delta\sigma $  at low energies ? If we try to blame them for the said difference, we will see that their contribution is only a small part of the visible effect.

It was understood by the authors of Ref.\cite{isrod} as they mentioned:
"{\it When we compare the available models to these data we find that none of them describes the data adequately.}"

That was true that no "adequate description" was provided that time but, nonetheless, the result did not remain unnoticed. In Ref.\cite{gaur} it was even called "the new great ISR discovery" with an "intriguing question":

"{\it Is it the maximal odderon growth?}"


\begin{wrapfigure}[43]{l}{95mm}
\vspace{2.1mm}
\includegraphics[width=95mm]{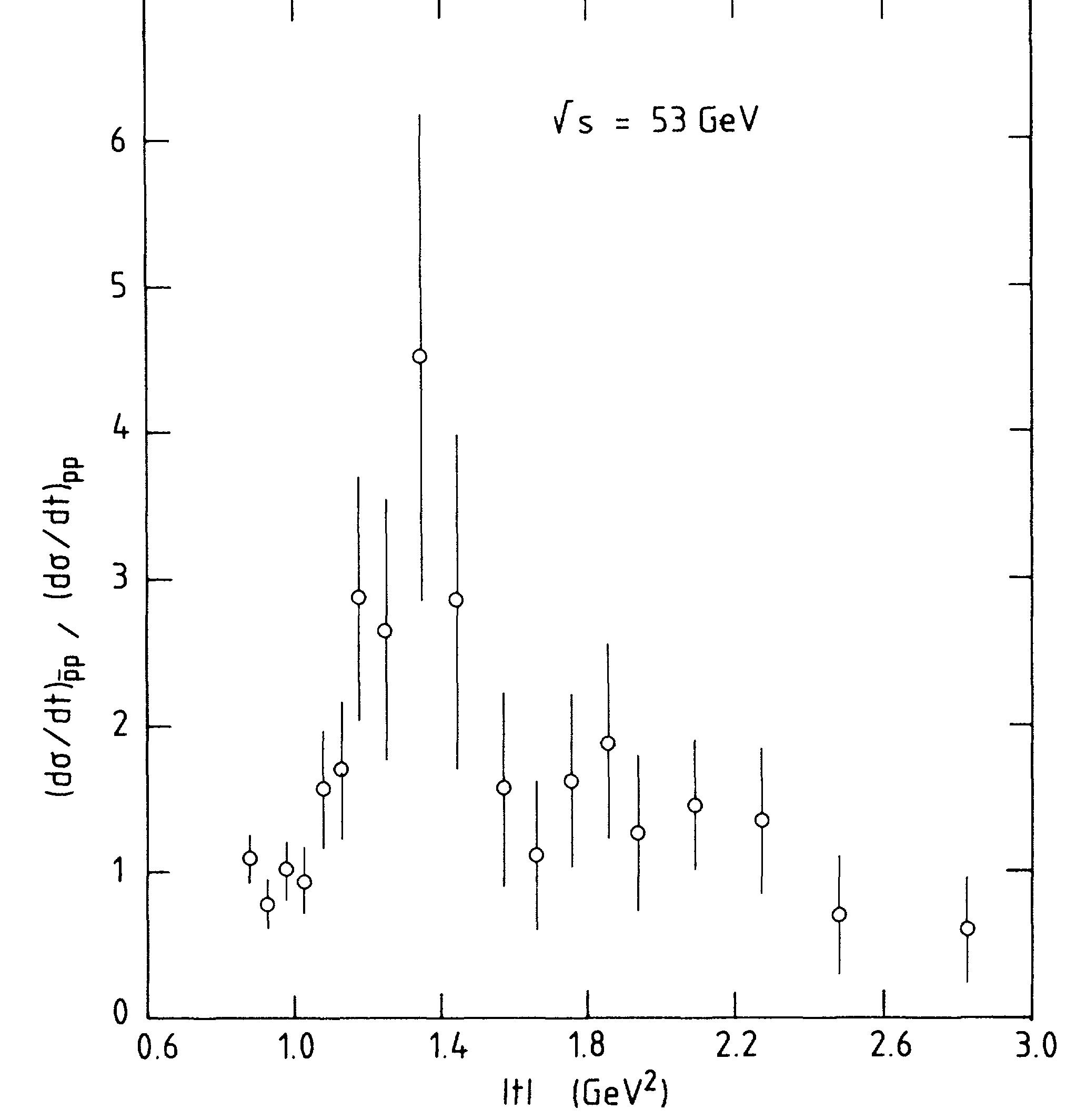}
\vspace{-6.6mm}
\caption{The ratio $d\sigma^{\bar{p}p}/d\sigma^{pp}$ at 53 GeV \cite{isrod}.}
\label{p5}
\vspace{2.1mm}
\includegraphics[width=95mm]{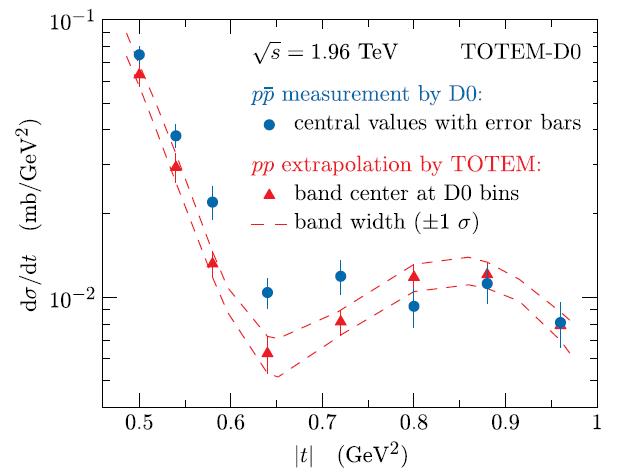}
\vspace{-6.6mm}
\caption{The $\bar{p}p$ and (extrapolated) $pp$ data at $\sqrt{s}=
1.96\mbox{ TeV}$ \cite{odde}.}
\label{p6}
\end{wrapfigure}

As we already mentioned the "Maximal Odderon", unfortunately,  was not acceptable on conceptual grounds. 

Howbeit, we have to admit that there was some new C-odd interaction agent observed in the experiment \cite{isrod}  which was not of pure quark origin as $ \rho, \omega$ etc. In other words that, in this blurry meaning, the Odderon was  discovered already 36 years ago.

But even if we admit this we do not know if this effect survives at high energies or dies off?

Meanwhile, the TOTEM Collaboration made measurements of the $ pp $ elastic scattering at 2.76 TeV. The closest results in energy was the DO (FNAL) measurement of the $\bar{p}p$ elastic scattering at 1.96 TeV. For lack of anything better, it was decided to compare the cross sections, albeit not at the same, but at relatively close energies.The comparison has shown that the effect persists, although less pronounced.

 Soon afterwards an attempt was  made\cite{odde}  with help of a specially designed extrapolation technique of the "data transfer" to provide the comparison at the same energy (1.96 TeV). The result appeared qualitatively the same with minor quantitative differences (Fig.6).

 Some qualitative estimate of the energy dependence can be made if to consider the ratio

$$d\sigma^{\bar{p}p}/d\sigma^{pp} = f(\sqrt{s}, \tau= \mid t\mid/\mid
t_{\mbox{dip}}(s)\mid).$$

The function $f( \sqrt{s} , \tau)$ seems to be 1 almost at all $\tau$
 except a "bell" in the vicinity of $\tau = 1$.

 Then we can obtain for the height of the bell that \[f(53\mbox{ GeV},1)-1
 = 3.5\pm 1.7 \]

 while

 \[f(1960\mbox{ GeV},1) -1 = 0.67\pm 1\sigma (?) .\]

 Unfortunately, the values of the extrapolated $ pp $ data at 1.96 TeV are still kept secret, so we could not estimate the errors better and make a picture like Fig.5.

 What can we conclude from this story?
\newpage

1. The discovery of the Odderon as a new C-odd force superior to the "old" C-odd forces from the secondary quark Reggeons was successfully confirmed in the energy interval $ 53 \div 2000 $ GeV.

2. The Odderon effect in the sense described above weakens with energy albeit very slow.

It remains to understand the Odderon nature in terms of, say, $ j $-plane singularity and to clarify its particle content.

We cannot agree that , as done in Ref.\cite{odde}, with a reference to the paper \cite{nicla}, that there was a "colorless C-odd gluonic compound" observed because the present data cannot inform us about quark-gluon content of the exchange. Only a direct detection of a state associated with this exchange with definition of its mass, width and spin-parity can be qualified as such an evidence. This can be compared with the discovery of pion occured only 12 years after publication of the Yukawa paper. Unfortunately, we can not  say that the paper \cite{nicla} with its 

erroneous theoretical content and bad description quality ($  p-value=8.5\cdot 10^{-71}$) \footnote{One can find the corresponding criticism in Ref.\cite{ptr1}. }  can be likened to the Yukawa paper.

At the same time, we would like to pay due tribute to the commendable tenacity of the main and pioneering proponent of the Odderon, Basarab Nicolescu, who devoted many years to enthusiastic promotion of this idea.

Comments: Presented at the Low-$x$ Workshop, Elba Island, Italy, September 27--October 1 2021.

\section*{Acknowledgements}

V.P. thanks Christophe Royon for invitation to give this talk at a very interesting Workshop.


\begin{thebibliography}{}





\bibitem{luk} 

L. Lukaszuk and B. Nicolescu,



Lett.Nuovo Cim. \textbf{8} (1973) 405-413;



Kyungsik Kang and Basarab Nicolescu,



Phys.Rev.\textbf{D 11} (1975) 2461.





\bibitem{amb}

M. Ambrosio et al.,



Phys.Lett., \textbf{B115} (1982)495.





\bibitem {Ow}



D. L. Owen,



AIP Conf.Proc. 98 (1983) 293-311.



\bibitem{ed}

V.V. Ezhela et al. 



NATO Sci.Ser.II 101 (2003) 47-6;

e-Print: hep-ph/0212398 [hep-ph].



\bibitem{Ant2}



TOTEM Collaboration, G. Antchev et al.,



Eur.Phys.J.\textbf{C 79} (2019) 2, 103.



\bibitem {ua4/1}

D. Bernard et al. UA4 Collaboration,



Phys. Lett. \textbf{B198}(1987) 583.

  

\bibitem {nic2}

Denis Bernard, Pierre Gauron and Basarab Nicolescu,



Phys.Lett.\textbf{B 199} (1987) 125.



\bibitem {ua4/2}



C. Augier et al. UA4/2 Collaboration,



Phys.Lett.\textbf{B 316} (1993) 448.



\bibitem{TOT1}

G. Antchev et al. TOTEM Collaboration,



Eur. Phys. J. \textbf{C 79} (2019) 785.



\bibitem{nic3}

E.Martynov and  B. Nicolescu, 



Phys. Lett. \textbf{B 778}, 414 (2018).



\bibitem {ptr1}

Vladimir A. Petrov, 



Eur.Phys.J.\textbf{C 81} (2021) 670.



\bibitem{isrod}

A. Breakstone et al.,



Phys.Rev.Lett.\textbf{54}(1985)2180.



\bibitem {gaur}



Pierre Gauron, Basarab Nicolescu, Elliot Leader,



Proceedings of EDS-85 (IPNO/TH 85-47) 26. 



\bibitem {odde}



TOTEM and D0 Collaborations,

V.M. Abazov et al.,



Phys.Rev.Lett. \textbf{127 }(2021) 6, 062003.



\bibitem {nicla}



Evgenij Martynov and Basarab Nicolescu,



Eur.Phys.J.\textbf{C 79} (2019) 6, 461.





\end{thebibliography}
\end{document}